\documentclass[apj]{emulateapj}
\slugcomment{{\sc Accepted for publication in the Astrophysical Journal:} April 20, 2010}
\usepackage{epsfig, natbib}
\usepackage{subfigure}

\begin{document}

\title{\textit{Swift}-BAT observations of the recently discovered magnetar SGR 0501+4516}
\author{Harsha S. Kumar\altaffilmark{1}, Alaa. I. Ibrahim\altaffilmark{2,3}
\& Samar Safi-Harb\altaffilmark{1,4}}
\altaffiltext{1}{Department of Physics \& Astronomy, University of Manitoba, Winnipeg, MB R3T 2N2, Canada; harsha@physics.umanitoba.ca}
\altaffiltext{2}{Department of Physics, American University in Cairo; Faculty of Science, Cairo University, Cairo, Egypt}
\altaffiltext{3}{Kavli Institute for Astrophysics and Space Research, Massachusetts Institute of Technology, Cambridge, MA 02139, USA; ai@space.mit.edu}
\altaffiltext{4}{Canada Research Chair; samar@physics.umanitoba.ca}

\begin{abstract}

We present results on the Soft Gamma Repeater
(SGR)~0501+4516, discovered by the \textit{Swift} Burst Alert Telescope (BAT) on 2008 August
22. More than 50 bursts were identified from this
source, out of which 18 bursts had enough counts to carry out spectral analysis.
We performed time-averaged spectral analysis on these 18 bursts using 8 models,
among which the cut-off powerlaw and the two-blackbody models provided the best fit
in the 15--150 keV energy range. 
The cut-off powerlaw model fit yields a mean photon index $\Gamma_{CPL}$ = 0.54$\pm$0.11 and a cut-off energy $E_C$ = 19.1$\pm$1.8 keV for the bursts. The mean hard and soft blackbody temperatures are found to be $kT_{BB_h}$ = 12.8$\pm$0.7 keV and $kT_{BB_s}$ = 4.6$\pm$0.5 keV, respectively, and are anti-correlated with the square of the radii of the hard and soft emitting regions ($R_{BB_h}$ and $R_{BB_s}$) as $R_{BB_h}^2$ $\propto$ $kT^{-5.8}$ and $R_{BB_s}^2$ $\propto$ $kT^{-2.7}$, respectively. The soft and hard component temperatures with different indices support the idea of two distinct emitting regions with the hard component corresponding to a smaller radius and the soft component corresponding to a larger radius, which further corroborate the idea of the propagation of extraordinary (E) and ordinary (O) mode photons across the photosphere, as predicted in the magnetar model. We notice strong burst fluence--duration correlation as well as hardness ratio--duration and hardness ratio--fluence anti-correlations for the SGR 0501+4516 bursts. The burst fluences range from $\sim$ 4.4$\times$10$^{-9}$ ergs~cm$^{-2}$ to $\sim$ 2.7$\times$10$^{-6}$ ergs~cm$^{-2}$, consistent with those observed for typical short SGR bursts.

\end{abstract}

\keywords{gamma rays: bursts --- stars: individual (SGR 0501$+$4516) --- stars: neutron --- X-rays: bursts}

\section{Introduction}
The \textit{Swift} mission (Gehrels et al. 2004), launched in November 2004 to explore the gamma-ray bursts (GRBs), has provided an excellent opportunity to detect and study the $\gamma$-ray activities from known Soft Gamma Repeaters (SGRs) as well as to discover new SGRs. SGRs, belonging to the class of `magnetars' (Duncan \& Thompson 1992), are highly magnetized ($B$ $\sim$ 10$^{14-15}$ G) and slowly rotating ($P$ $\sim$ 2$-$12 s) neutron stars characterized by short, bright bursts of hard X-rays and soft $\gamma$-rays (see Mereghetti 2008 for a recent review). During its active state, an SGR can go through periods of intense bursting activity lasting from a few days to months, however, it can also remain dormant for many years. The bursts, often varying in duration, are classified into three main categories: short, intermediate and giant flares.
The short bursts are the typical kind observed during an outburst, marked by timescales $\sim$ 0.1--0.5~s and luminosities $\sim$ 10$^{38}$--10$^{41}$ ergs s$^{-1}$, whereas the intermediate bursts have timescales $\sim$ 1--60~s and luminosities $\sim$ 10$^{41}$--10$^{43}$~ergs~s$^{-1}$ (Mereghetti 2008). Giant flares, the rare and unique events, are distinguished by their extreme energies (10$^{44}$--10$^{47}$~ergs~s$^{-1}$), long durations ($\sim$ hundreds of seconds), and the presence of a coherent pulsating decaying tail, consistent with the spin period of the neutron star. Persistent X-ray emission, interpreted as originating due to the magnetospheric currents driven by twists in the evolving ultra-high magnetic field (Thompson \& Duncan 1995), has also been observed from SGRs in the 0.1--10 keV band with a typical X-ray luminosity of $\sim$ 10$^{35}$ ergs~s$^{-1}$, and the spectrum is generally described by an absorbed powerlaw (PL) plus a blackbody component (Mereghetti 2008).

According to the magnetar model (Thompson \& Duncan 1995), the dominant form of energy powering an SGR is its decaying ultra-strong magnetic field. The surface of the neutron star is heated and fractured by instabilities generating Alfv\'{e}n waves that accelerate electrons, and in turn give away their energy in short bursts. The model also suggests that the high-energy dissipated remains trapped in the magnetosphere as the `trapped fireball', and it shrinks in size with time. Alternatively, the bursts can arise from heating of the corona by magnetic reconnection in the stellar magnetosphere
resulting in intermediate-type flares (Lyubarsky 2002). The giant flares are caused by the sudden rearrangement of the star's magnetic field producing global crustal fractures (Thompson \& Duncan 1995).

\begin{table*}[th]
\caption{Summary of SGR 0501+4516 bursts observed with \textit{Swift-BAT}.}
\begin{tabular}{l l l l l l}
\hline
Triggers  & Date of Observation & Exposure$^{a}$  & Bursts & T$_{100}$  & Count-rate$^b$ \\
& & (ks) & & (s) & (counts s$^{-1}$) \\
\hline

00321174000 & 2008 August 22 & 34.8 & 321174A  & 0.132 & 1.22$\pm$0.06 \\
& & &  321174B & 0.036 & 0.58$\pm$0.07 \\
\hline
00321177000 & 2008 August 22 & 3.0 & 321177  & 0.150 & 1.33$\pm$0.06  \\
\hline
00321174001 & 2008 August 22 & 12.8 & 321174001 & 0.200 & 0.48$\pm$0.03 \\
\hline
00321252000 & 2008 August 22 & 2.6 & 321252  & 0.078 & 0.43$\pm$0.04 \\
\hline
00321481000 & 2008 August 23 & 3.2 & 321481A  & 0.419 & 7.35$\pm$0.21 \\
& & &  321481B  & 0.263 & 1.75$\pm$0.05\\
\hline
00321551000 & 2008 August 23 & 4.2 & 321551  & 0.487 & 15.2$\pm$0.40\\ \hline
00321574000 & 2008 August 23 & 3.4 & 321574  & 0.044 & 0.44$\pm$0.06\\ \hline
00321583000 & 2008 August 23 & 4.0 & 321583A & 0.043 & 0.61$\pm$0.06\\
& & & 321583B  & 0.068 & 0.22$\pm$0.04\\
& & & 321583C  & 0.092 & 0.43$\pm$0.04\\
&& &  321583D  & 0.119 & 0.62$\pm$0.04\\
& & & 321583E  & 0.332 & 2.38$\pm$0.06\\
& & & 321583F  & 0.140 & 1.11$\pm$0.04\\
& & & 321583G  & 0.059 & 0.33$\pm$0.04\\ \hline
00323192000 & 2008 September 01 & 3.3 & 323192  & 0.044 & 0.33$\pm$0.05\\ \hline
00323650000 & 2008 September 03 & 4.3 & 323650  & 0.151 & 0.48$\pm$0.06\\
\hline
\end{tabular}
\tablecomments{T100 is the total burst duration obtained by noting the start and the end time of the burst, manually. See \S\ref{2} for details. \\
$^{a}$ Total BAT exposure for the trigger numbers. \\
$^b$ Count-rates are in the 15--150 keV energy range.}
\end{table*}

SGR~0501+4516 is a recently discovered SGR by the \textit{Swift} $\gamma$-ray observatory (Holland et al. 2008; Barthelmy et al. 2008). \textit{Rossi X-ray Timing Explorer} (RXTE) observations discovered a 5.7~s X-ray pulsating counterpart (Gogus et al. 2008),  with a dipole magnetic field $B$ = 3$\times$10$^{14}$ G  estimated from its period and spin-down rate (Woods et al. 2008), thus confirming the magnetar nature of the source. Soon after, the GLAST Burst Monitor (GBM) onboard \textit{Fermi} satellite triggered and located the bursts from SGR 0501+4516 (van der Horst \& Connaughton 2008; McBreen et al. 2008).  Multiwavelength observations reported the detection of its infrared (Tanvir et al. 2008; Rea et al. 2008) and optical (Fatkhullin et al. 2008; Rol et al. 2008) counterparts. However, no radio emission was detected from this source during the outburst (Kulkarni \& Frail 2008; Gelfand et al. 2008). $Suzaku$ observations of the SGR performed from 2008
August 26--27 detected 32 bursts with the X-ray imaging spectrometer (XIS) and the hard X-ray detector (HXD). The persistent X-ray  emission obtained with the XIS in the 0.4--10 keV range was best fitted by the combination of a blackbody ($kT$ = 0.69$\pm$0.01 keV) plus  a PL ($\Gamma$ = 2.8$\pm$0.1) component, modified by an interstellar absorption $N_H$ = (0.89$\pm$0.08)$\times$10$^{22}$~cm$^{-2}$ (Enoto et al. 2009). The 1--200 keV combined XIS+HXD burst spectrum was described by a two-component blackbody model having soft and hard temperatures of  $kT$ = 3.3$_{-0.4}^{+0.5}$ keV and 15.1$_{-1.9}^{+2.5}$ keV,  respectively, and was interpreted as the  population of ordinary (O) and extraordinary (E) mode photons propagating across the magnetosphere (Enoto et al. 2009).  \textit{Konus-Wind} $\gamma$-ray burst spectrometer observed the SGR bursts from 2008 August 23--26 and the 20--200 keV spectra  were best fitted by either a powerlaw with an exponential cut-off (CPL) model or an optically-thin thermal bremsstrahlung model (OTTB), both of  which gave a peak energy ($E_p$) or temperature ($kT$) in the range of 20--45 keV (Aptekar et al. 2009). Subsequently, five \textit{XMM-Newton} and two \textit{INTEGRAL} observations of the source were reported, with a hard X-ray variable source detected by \textit{INTEGRAL} only during the first pointing (Rea et al. 2009).  The phase-coherent timing analysis performed with \textit{XMM-Newton}, \textit{Suzaku}-XIS and \textit{Swift}-XRT data further refined the period $P$ = 5.7620695(1)~s, period derivative $\dot{P}$ = 6.7(1)$\times$10$^{-12}$~s~s$^{-1}$, and  magnetic field $B$ $\sim$ 7$\times$10$^{13}$ -- 2$\times$10$^{14}$ G, with the evidence of a second period derivative  $\ddot{P}$ = $-$1.6(4)$\times$10$^{-19}$ s~s$^{-2}$ (Rea et al. 2009). Moreover, the spectrum of the outburst indicates a  trend of spectral softening, with the blackbody component decaying slower than the PL component and the spectral evolution changing  from phase to phase (Rea et al. 2009).

In this paper, we report the \textit{Swift}-BAT observations of the recently discovered magnetar SGR~0501+4516. The paper is organized as
follows: \S\ref{2} and \S\ref{3} describe the observations and the burst spectroscopic analysis, respectively. In \S\ref{4}, we present the results of the time-averaged spectral and statistical analyses of the
SGR bursts, which are then discussed in \S\ref{5}. The conclusions are summarized in \S\ref{6}.

\section{Observations and data reduction}
\label{2}

The initial outburst from SGR 0501+4516 was discovered with the Burst Alert Telescope (BAT) on 2008 August 22 (BAT calculated R.A = 05h 01m 04s, Dec = +45d 16\hbox{$^{\prime}$} 20\hbox{$^{\prime\prime}$}, with an uncertainty of 3\hbox{$^{\prime}$} radius; Holland et al. 2008; Barthelmy et al. 2008). \textit{Swift}-BAT is a highly sensitive and large field of view (1.4~sr half-coded) hard X-ray telescope using a coded aperture mask operating in the 15--350 keV energy range (Barthelmy et al. 2005). The BAT detector plane is made of 32,768 pieces of CdZnTe (CZT: 4 $\times$ 4 $\times$ 2~mm) and the coded-aperture mask consists of $\sim$ 52,000 lead tiles (5 $\times$ 5 $\times$ 1~mm) having a 1-m separation between mask and detector plane (Barthelmy et al. 2005).  We used data from all the BAT triggers (see Table~1) and observation sequences (from \textit{00321174001 to 00321174061}) available for this source. Intense bursting activity was observed between 2008 August 22--23 with the last BAT trigger on 2008 September 3. The standard BAT software distributed within FTOOLS under the HEASoft package\footnote{\tiny{http://heasarc.gsfc.nasa.gov/lheasoft/}} (version 6.4.1) and the latest calibration files available were used to analyze the data. The burst pipeline script, \textit{batgrbproduct\footnote{\tiny{http://heasarc.nasa.gov/docs/swift/analysis/threads/batgrbproductthread.html}}}, was run to process the BAT trigger events. In this study, we screened out the faint bursts (characterized by a count-rate $\lesssim$ 0.1 counts s$^{-1}$) from the $\sim$ 50 total bursts observed, leaving 18 bursts for the temporal and spectral analysis. In the cases where more than one burst was observed in a trigger, the bursts were denoted by the letters `A', `B', etc., preceded by the trigger number (for example, the two bursts considered for spectral analysis from the trigger $000321174000$ were named as $321174A$ and $321174B$). In Table 1, we present the summary of the bursts identified for spectral and temporal analysis along with their total BAT exposure and count-rates (counts s$^{-1}$) in the 15--150 keV energy range.


Background-subtracted light curves were made for all the bursts using the task \textit{batbinevt} in the 15--25, 25--50, 50--100 and 100--150 keV energy ranges binned at 4 ms time resolution after applying the maskweighting technique using the \textit{batmaskwtevt} task. Maskweighting\footnote{\tiny{http://heasarc.nasa.gov/docs/swift/analysis/threads/batmaskwtthread.html}} is defined as the background-subtracted counts per fully illuminated detector for an equivalent on-axis source and involves assigning each event a weight according to the illumination fraction of the detector it was detected in. A sample of the burst light curves is displayed in Figure 1. To study the burst durations, we determined $T_{90}$ from the light curves by running the task \textit{battblocks}, where $T_{90}$, the standard parameter describing the burst duration of GRBs,  is defined as the time to accumulate 5$\%$ to 95$\%$ of the observed photons. However, \textit{battblocks} failed to determine the burst's duration for a few weak bursts. Hence, $T_{100}$ interval was used in our analysis, which was determined by noting the start and the end time of the burst emission, manually, and  the $T_{100}$ values range from $\sim$ 0.03~s to $\sim$ 0.5~s with an average value of 0.16$\pm$0.02~s. In Figure 2, we show the $T_{100}$ duration histogram of SGR 0501+4516's bursts, overplotted with the $T_{90}$ duration histogram ranging from 0.07~s to 0.25~s, for comparison. The solid and the dotted curves are the lognormal Gaussian best-fits, peaking at 0.06$\pm$0.02~s and 0.09$\pm$0.02~s with a Gaussian width $\sigma$ of 0.8$\pm$0.1~s and 0.5$\pm$0.1~s for $T_{100}$ and $T_{90}$, respectively.

Response matrices for BAT spectra were generated for each of the 18 bursts using the \textit{batdrmgen} task and the spectral fitting was restricted between 15--150 keV band, since the BAT mask becomes transparent around 150 keV. Finally, a systematic error was applied to the BAT spectra using the \textit{batphasyserr} task in order to account for residuals in the response matrix. The spectral analysis was performed using XSPEC v12.4.1. Errors quoted are at the 90\% confidence level.

\section{Burst spectroscopy}
\label{3}

The time-averaged spectral analysis was performed using the $T_{100}$ interval. The bursts spectra were subsequently fitted with 4 single component models: PL, CPL (characterized by an exponential cut-off energy $E_C$ and a powerlaw photon index $\Gamma_{CPL}$), thermal bremsstrahlung (Bremss), blackbody radiation with normalization proportional to the surface area (Bbodyrad in XSPEC; abbreviated as BB throughout this paper), and 4 double models by adding a Bbodyrad component to all the above mentioned single models: BB+PL, BB+CPL, BB+Bremss and BB+BB. We first averaged the measured total $\chi^2$ for all the bursts to determine their average reduced $\chi^2$ ($\langle\chi^2_{\nu}\rangle$, where $\nu$ is the number of degrees of freedom) and their \textit{F-test} probability. The results, considered to give a qualitative analysis of the spectral model fit goodness, are summarized in Table~2 in the order of increasing $\langle\chi^2_{\nu}\rangle$. Among the single-component models, the CPL model gave the best fit ($\langle\chi^2_{\nu}\rangle$ = 1.004) and a single PL model gave the worst-fit ($\langle\chi^2_{\nu}\rangle$ = 2.766). The Bremss model, although considered best to describe the hard X-ray spectra of SGRs, also gave a large $\langle\chi^2_{\nu}\rangle$ and failed to reproduce the spectral energy distribution for most of the bursts in the high-energy band, as reported by various authors in their study of other SGRs (for example, Feroci et al. 2004; Israel et al. 2008).


\begin{table}
\caption{Summary of the average $\chi^2$ and $F$-test values.}
\begin{tabular}{l l l l l l}
\hline
Model & $\langle\chi^2\rangle$ & $\nu$ & $\chi^2_{\nu}$ & $F$-test & $F$-test\\
&  & &  & value & probability \\
\hline
CPL & 1012.45 & 1008 & 1.004 & & \\
Bremss & 1448.19 & 1026 & 1.411 &  \\
BB & 1519.46 & 1026 & 1.481 & \\
PL & 2837.56 & 1026 & 2.766  \\
\hline
BB+BB & 1003.13 & 990 & 1.013 & 14.155 & 1.54$\times$10$^{-66}$ \\
BB+CPL & 989.34 & 972 & 1.018  & 0.6307 & 0.96 \\
BB+PL & 1300.77 & 990 & 1.314 & 32.490 & 1.57$\times$10$^{-141}$ \\
BB+Bremss & 1331.23 & 990 & 1.345 & 2.4161 & 8.21$\times$10$^{-6}$ \\
 \hline
\end{tabular}
\tablecomments{Obtained for the 18 time-averaged burst spectra of SGR 0501+4516 using \textit{Swift-BAT}. See \S\ref{3} for details.}
\end{table}


We next investigated the significance of the double models: BB+PL, BB+CPL, BB+Bremss and BB+BB. As shown in Table~2, the BB+BB and BB+CPL models provide comparable $\langle\chi^2_{\nu}\rangle$ values, but the BB+CPL model was ruled out  because of the large \textit{F-test} probability. For the BB+Bremss and BB+PL models, even though the \textit{F-test} probability suggests significant improvement of the fit by the addition of a BB component, the $\langle\chi^2_{\nu}\rangle$ values obtained are relatively high. Therefore, out of the 8 trial models, the CPL and BB+BB models yielded statistically acceptable $\chi^2_{\nu}$ values.  Figures 3 (a) and (b) show a sample of the best-fit spectra obtained with the CPL and BB+BB models for a burst (321583E) of duration $\sim$ 0.33~s.

\section{Results}
\label{4}

In the following subsections, we present the results from our spectroscopic analysis and also investigate the correlation between the spectral and temporal properties of the observed bursts from SGR 0501+4516.

\subsection{Time-averaged spectral parameters}
\label{4.1}

The best-fit time-averaged spectral parameters of the bursts are summarized in Table~3. We have determined the cooler and hotter BB radii ($R_{BB_s}$ and $R_{BB_h}$) from the two-component BB model normalization $K$ $\sim$ $R^2_{km}$/$D^2_{10}$, where $R_{km}$ is the radius of the emitting region in km and $D_{10}$ is the distance to the source in units of 10 kpc. Although the exact distance to SGR 0501+4516 is not yet known, a distance of 1.5 kpc based on the proximity of its direction to the supernova remnant HB9 has been suggested by Gaensler \& Chatterjee (2008). More recent studies used distance estimates of 1.5~kpc (Aptekar et al. 2009), 5~kpc (Rea et al. 2009) and 10~kpc (Enoto et al. 2009). In this work, we assume a distance to the source in units of 10 kpc ($D_{10}$) to be consistent with the XSPEC BB model normalization and scale all the derived parameters in units of $D_{10}$.   We determined a mean radius of $\sim$ 7.2$\pm$1.5 $D_{10}$~km and $\sim$ 0.9$\pm$0.2 $D_{10}$~km for the soft and hard BB components, respectively, which are consistent with the cooler ($\sim$ 8.9$^{+2.9}_{-2.1}$ $D_{10}$ km) and  hotter ($\sim$ 0.46$^{+0.16}_{-0.14}$ $D_{10}$ km) BB radii obtained for SGR 0501+4516 using \textit{Suzaku} observations (Enoto et al. 2009). The histograms of the best-fit spectral parameters ($\Gamma_{CPL}$, $E_C$, $kT_{BB_{s}}$, $kT_{BB_{h}}$) and the BB radii of the soft and hard components ($R_{BB_{s}}$ and $R_{BB_{h}}$) are shown in Figure~4. In Table~4, we summarize the mean value obtained from a Gaussian fit to the histogram plots of all spectral parameters and the inferred BB radii (see Figure~4). The results obtained for SGR 0501+4516 are in good agreement with those obtained for the typical short-duration bursts observed in other SGRs such as SGR~1900+14 and SGR~1806$-$20 (Olive et al. 2004; Feroci et al. 2004; Nakagawa et al. 2007; Israel et al. 2008).

\begin{table*}[th]
\caption{Summary of the spectral fit results to the SGR 0501+4516 bursts}
\begin{tabular}{l l l l l l l l l l}
\hline
Bursts  & $\Gamma_{CPL}$ & $E_C$ & $kT_{BB_h}$ & $R_{BB_h}$ & $kT_{BB_s}$ & $R_{BB_s}$ & Fluence & Flux & Luminosity\\
&  & (keV) & (keV) & (km) & (keV) & (km) & (ergs cm$^{-2}$) & (ergs cm$^{-2}$ s$^{-1}$) & (ergs s$^{-1}$) \\
\hline
321174A  & -1.9$_{-1.4}^{+1.2}$ & 9.1$_{-1.4}^{+3.8}$  & \nodata & \nodata & \nodata & \nodata & 6.7$^{+0.1}_{-0.1}$$\times$10$^{-8}$ & 5.1$^{+0.3}_{-0.6}$$\times$10$^{-7}$ & 6.1$^{+0.4}_{-0.8}$$\times$10$^{39}$ \\

321174B  & -0.5$_{-2.6}^{+2.0}$ & 14.7$_{-8.7}^{+56.5}$ & 10.7$_{-1.9}^{+0.4}$ & 1.2$_{-0.6}^{+0.9}$ & 3.5$_{-3.5}^{+2.2}$ & 5.9$_{-2.9}^{+3.9}$ & 7.0$^{+0.0}_{-0.1}$$\times$10$^{-9}$ & 1.9$^{+0.2}_{-1.9}$$\times$10$^{-7}$ & 2.3$^{+0.2}_{-2.3}$$\times$10$^{39}$\\

321174001 & 1.4$_{-0.9}^{+0.7}$ & 56.3$_{-13.5}^{+20.4}$ & 16.8$_{-4.0}^{+8.0}$ & 0.4$_{-0.2}^{+0.5}$ & 5.4$_{-1.5}^{+1.7}$ & 3.5$_{-1.0}^{+4.0}$ & 2.4$_{-0.2}^{+0.1}$$\times$10$^{-8}$ & 2.0$_{-1.6}^{+0.1}$$\times$10$^{-7}$ & 2.4$^{+0.1}_{-1.9}$$\times$10$^{39}$\\

321177 & 1.0$_{-0.6}^{+0.6}$ & 25.9$_{-8.0}^{+15.6}$ & 11.4$_{-1.2}^{+1.6}$ & 1.5$_{-0.3}^{+0.5}$ & 4.0$_{-0.8}^{+0.9}$ & 11.0$_{-0.8}^{+0.8}$ & 7.7$^{+0.0}_{-0.4}$$\times$10$^{-8}$ & 5.1$^{+0.2}_{-2.8}$$\times$10$^{-7}$ & 6.1$^{+0.2}_{-3.3}$$\times$10$^{39}$\\

321252 & -1.9$_{-2.1}^{+1.5}$ & 10.9$_{-2.1}^{+5.3}$ & 11.3$_{-1.2}^{+1.1}$ & 1.0$_{-0.2}^{+0.2}$ & 2.4$_{-2.4}^{+0.3}$ & 12.9$_{-2.3}^{+2.3}$ & 1.3$^{+0.0}_{-0.1}$$\times$10$^{-8}$ & 1.6$^{+0.1}_{-0.6}$$\times$10$^{-7}$ & 1.9$^{+0.2}_{-0.7}$$\times$10$^{39}$\\

321481A & 0.9$_{-0.3}^{+0.3}$ & 20.6$_{-2.6}^{+3.1}$ & 11.2$_{-0.7}^{+1.0}$ & 3.1$_{-0.6}^{+0.8}$ & 5.1$_{-0.6}^{+0.7}$ & 16.4$_{-0.4}^{+0.4}$  & 1.1$^{+0.0}_{-0.1}$$\times$10$^{-6}$ & 2.7$^{+0.0}_{-0.3}$$\times$10$^{-6}$ & 3.2$^{+0.0}_{-0.4}$$\times$10$^{40}$ \\

321481B & 0.1$_{-0.6}^{+0.5}$ & 21.1$_{-4.2}^{+5.5}$ & 13.6$_{-2.3}^{+7.7}$ & 1.2$_{-0.5}^{+1.0}$  & 6.6$_{-2.6}^{+2.3}$ &  4.1$_{-0.9}^{+7.6}$  & 1.9$^{+0.0}_{-1.4}$$\times$10$^{-7}$ & 7.4$^{+0.1}_{-5.4}$$\times$10$^{-7}$ & 8.8$^{+0.1}_{-6.5}$$\times$10$^{39}$ \\

321551 & 0.5$_{-0.2}^{+0.2}$ & 18.4$_{-1.4}^{+1.6}$ & 13.1$_{-1.0}^{+1.3}$ &  2.7$_{-0.6}^{+1.0}$  & 6.7$_{-0.6}^{+0.5}$ &  14.2$_{-1.4}^{+2.3}$  & 2.7$^{+0.0}_{-02}$$\times$10$^{-6}$ & 5.6$^{+0.1}_{-0.5}$$\times$10$^{-6}$ & 6.7$^{+0.1}_{-0.6}$$\times$10$^{40}$ \\

321574 &  0.8$_{-1.8}^{+0.9}$ & 50.9$_{-35.7}^{+97.9}$ & 25.8$_{-12.6}^{+3.3}$ & 0.2$_{-0.1}^{+0.1}$  & 8.0$_{-3.9}^{+4.5}$ & 1.7$_{-0.6}^{+5.9}$  &  1.0$^{+0.1}_{-0.1}$$\times$10$^{-8}$ & 2.3$^{+1.0}_{-2.3}$$\times$10$^{-7}$ & 2.7$^{+1.2}_{-2.7}$$\times$10$^{39}$ \\

321583A &  0.1$_{-2.4}^{+1.0}$ & 21.5$_{-12.3}^{+21.3}$ & 11.3$_{-2.3}^{+0.7}$ &  1.2$_{-0.3}^{+0.9}$  & 3.7$_{-1.1}^{+0.5}$ & 5.7$_{-2.5}^{+2.5}$  &  1.0$^{+0.0}_{-0.1}$$\times$10$^{-8}$ & 2.4$^{+0.2}_{-2.4}$$\times$10$^{-7}$ & 2.9$^{+0.2}_{-2.9}$$\times$10$^{39}$ \\

321583B & 0.2$_{-2.4}^{+1.6}$ & 34.4$_{-7.0}^{+6.9}$ & 19.1$_{-2.9}^{+2.2}$ &  0.2$_{-0.1}^{+1.0}$  &  6.7$_{-1.7}^{+0.9}$ &  1.2$_{-0.4}^{+0.4}$  &  7.0$^{+0.0}_{-0.1}$$\times$10$^{-9}$ & 1.0$^{+0.0}_{-1.0}$$\times$10$^{-7}$ & 1.2$^{+0.0}_{-1.2}$$\times$10$^{39}$ \\

321583C & 0.6$_{-1.4}^{+1.1}$ & 42.9$_{-10.8}^{+15.6}$ & 15.5$_{-2.9}^{+5.1}$ &  0.5$_{-0.2}^{+0.2}$  &  3.7$_{-1.6}^{+3.9}$ &  5.4$_{-2.6}^{+2.6}$  &  1.8$^{+0.0}_{-0.1}$$\times$10$^{-8}$ & 1.9$^{+0.1}_{-1.3}$$\times$10$^{-7}$ & 2.3$^{+0.1}_{-1.5}$$\times$10$^{39}$\\

321583D & -0.5$_{-1.0}^{+0.8}$ & 23.9$_{-8.4}^{+17.3}$ & 15.8$_{-2.0}^{+4.8}$ &  0.7$_{-0.3}^{+0.3}$  & 5.1$_{-2.9}^{+2.6}$ &  2.3$_{-0.7}^{+0.7}$  & 3.8$^{+0.0}_{-0.2}$$\times$10$^{-8}$ & 3.2$^{+0.3}_{-2.1}$$\times$10$^{-7}$ & 3.8$^{+0.3}_{-2.5}$$\times$10$^{39}$ \\

321583E  & -0.7$_{-0.4}^{+0.3}$ & 18.6$_{-2.3}^{+2.7}$ & 13.4$_{-0.5}^{+0.5}$ & 1.8$_{-0.1}^{+0.1}$ & 3.3$_{-0.8}^{+1.2}$ & 14.8$_{-1.8}^{+1.8}$  &  3.7$^{+0.0}_{-0.1}$$\times$10$^{-7}$ & 1.1$^{+0.0}_{-0.2}$$\times$10$^{-6}$ & 1.3$^{+0.1}_{-0.2}$$\times$10$^{40}$\\

321583F  & -2.5$_{-5.5}^{+2.5}$ & 11.1$_{-0.4}^{+0.4}$ & \nodata & \nodata &  \nodata & \nodata & 7.2$^{+0.1}_{-0.7}$$\times$10$^{-8}$ & 5.1$^{+0.5}_{-5.1}$$\times$10$^{-7}$ & 6.1$^{+0.6}_{-6.1}$$\times$10$^{39}$ \\

321583G & 0.4$_{-3.3}^{+1.8}$ & 22.2$_{-6.2}^{+9.7}$ & \nodata & \nodata & \nodata & \nodata  & 3.6$^{+0.0}_{-0.4}$$\times$10$^{-8}$ & 1.2$^{+0.3}_{-1.2}$$\times$10$^{-8}$ & 1.5$^{+0.4}_{-1.5}$$\times$10$^{38}$ \\

323192  & -2.9$_{-0.2}^{+0.3}$ & 6.2$_{-1.0}^{+1.1}$ & \nodata & \nodata &\nodata & \nodata & 4.4$^{+0.1}_{-0.4}$$\times$10$^{-9}$ & 9.9$_{-8.2}^{+2.0}$$\times$10$^{-8}$ & 1.2$^{+0.2}_{-1.0}$$\times$10$^{39}$  \\

323650 & 0.2$_{-0.6}^{+0.5}$ & 20.6$_{-4.5}^{+6.2}$ & 12.2$_{-1.4}^{+2.5}$ &  1.5$_{-0.5}^{+0.8}$  &  4.6$_{-2.0}^{+2.3}$ &  7.1$_{-0.6}^{+0.6}$  & 9.7$^{+0.0}_{-1.0}$$\times$10$^{-8}$ & 6.4$_{-6.4}^{+0.1}$$\times$10$^{-7}$ & 7.7$^{+0.1}_{-7.7}$$\times$10$^{39}$\\

\hline
\end{tabular}
\tablecomments{The radius, fluence, flux and luminosity values quoted are in the 15--150 keV energy range, assuming a distance to the source in units of 10 kpc ($D_{10}$).
Errors quoted are at the 90 $\%$ confidence level. See \S\ref{4} for details. }
\end{table*}

The burst fluence, tabulated in Table~3, was estimated by multiplying the time-averaged flux (in units of ergs~cm$^{-2}$~s$^{-1}$ obtained from the spectral fits) by its respective burst duration $T_{100}$. For SGR 0501+4516, the fluence ranges from $\sim$ 4.4$\times$10$^{-9}$ ergs cm$^{-2}$ to $\sim$ 2.7$\times$10$^{-6}$ ergs cm$^{-2}$.  Observational results suggest that SGR bursts fluence ranges from $\sim$10$^{-10}$ -- $\sim$ 10$^{-4}$ ergs~cm$^{-2}$ and follows a powerlaw distribution (Mereghetti 2008). The corresponding burst energies are in the range of $\sim$ 4.2$\times$10$^{36}$ ergs to $\sim$ 2.8$\times$10$^{39}$ ergs. Assuming isotropic emission, we calculate the burst peak luminosity as $L$ = 4$\pi$$D^2_{10}$$F$, where $F$ is the time-averaged flux of the bursts (see Table~3), and find that it ranges from $\sim$ 1.5$^{+0.4}_{-1.5}$$\times$10$^{38}$ $D^2_{10}$~ergs s$^{-1}$ to $\sim$ 6.7$^{+0.1}_{-0.6}$$\times$10$^{40}$ $D^2_{10}$~ergs s$^{-1}$. We have also determined the luminosities corresponding to the soft and hard components of the BB+BB model; these are reported and discussed in \S\ref{5}.



\begin{table*}[th]
\caption{Summary of the Gaussian fit to the spectral parameters.}
\begin{tabular}{l l l l l l l l}
\hline
Parameters & $\Gamma_{CPL}$ & E$_C$ & $kT_{BB_h}$ & $kT_{BB_s}$  & $R_{BB_h}$ & $R_{BB_s}$ \\
 & & (keV) & (keV) & (keV) & (km) & (km) \\
\hline
Gaussian Mean of parameters & 0.54$\pm$0.11 & 19.1$\pm$1.8 & 12.8$\pm$0.7 & 4.6$\pm$0.5 & 0.9$\pm$0.2 & 7.2$\pm$1.5\\
Width of Gaussian distribution ($\sigma$) & 0.38$\pm$0.11 & 6.7$\pm$1.8 & 1.2$\pm$0.9 & 1.7$\pm$0.6 & 0.6$\pm$0.2 & 5.9$\pm$1.5 \\
 \hline
\end{tabular}
\end{table*}

\subsection{Statistical analysis of SGR 0501+4516}
\label{4.2}

Statistical studies have unveiled some basic properties such as the burst energy injection and radiation mechanisms of the SGRs (Gogus et al. 1999, 2000, 2001). In this subsection, we consider the statistical properties of the SGR 0501+4516 bursts. First, we investigated the fluence distribution of the SGR bursts with duration and estimated the significance of this correlation using the Spearman test (Spearman 1904).  The fluences show a strong positive correlation with $T_{100}$, with a PL index of 2.0$\pm$0.2 as shown in Figure 5. The Spearman rank order test applied to SGR 0501+4516 burst fluences and durations yielded a correlation coefficient $\rho$ = 0.9 and the probability ($P$) that this correlation is due to a random fluctuation is 3.0$\times$10$^{-7}$ corresponding to 95$\%$ confidence level. Such a correlation has been observed by Gogus et al. (2001) for SGR 1806-20 and SGR 1900+14, which was interpreted to be similar to the PL relation between the total energy and duration of the earthquakes established by Gutenberg \& Richter (1956).


Next, we examined the burst spectral variations with temporal properties to draw comparison to other SGRs. For that, we computed the hardness ratio, defined as HR = ($H$-$S$)/($H+S$), where $H$ and $S$ are the background subtracted hard and soft photon counts in the 25--150 keV and 15--25 keV energy bands, respectively.   Figures~6 (a) and (b) show HR plotted against the burst duration and fluence, respectively. We clearly see an anti-correlation in both cases and a powerlaw fitted to the data gave indices of $-$0.20$\pm$0.03 and $-$0.10$\pm$0.02, respectively. This anti-correlation was further quantified by carrying out the Spearman test and we obtain a correlation coefficient of $\rho$ = $-$0.7 ($P$ = 6.5$\times$10$^{-4}$) for HR vs. burst durations and $\rho$ = $-$0.8 ($P$ = 1.8$\times$10$^{-4}$) for HR vs. burst fluences.  We discuss the observed correlations in \S\ref{5}; a detailed analysis on the spectral evolution in SGR 0501+4516 will be presented elsewhere.


\section{Discussion}
\label{5}

We have carried out a comprehensive study of the spectral and temporal properties of the recently discovered SGR 0501+4516. The observed burst durations in SGR 0501+4516 typically range from $\sim$ 0.03--0.5~s and follow a lognormal distribution as seen in the case of other SGRs (for example, Gogus et al. 2001; Woods et al. 1999). $T_{90}$ durations of $\sim$ 0.093~s and $\sim$ 0.162~s were measured for SGRs 1806-20 and 1900+14 (Gogus et al. 2001), respectively, and $\sim$ 0.099~s for the anomalous X-ray pulsar (AXP) 1E 2259+486 (Gavriil et al. 2004). The $T_{100}$ duration of the bursts observed from SGR 0501+4516 showed a lognormal distribution peaking at $\sim$ 0.06$\pm$0.02~s with an average value of $\sim$ 0.16$\pm$0.02~s for the 18 bursts.
The short bursts observed from SGR 0501+4516 using \textit{Swift}-BAT were fitted with 8 spectral models, out of which the single-component CPL and the two-component BB+BB  models provided the best fit. Feroci et al. (1994) also suggested CPL and BB+BB as best-fit models for 10 short bursts from SGR 1900+14 observed using \textit{BeppoSAX} in the 1.5--100 keV energy range. We have also explored the possible correlations between the spectral and the temporal burst properties in SGR 0501+4516 and the implications of these observed features are discussed below in the context of the magnetar model (Thompson \& Duncan 1995).

As shown in \S\ref{4.2}, the SGR 0501+4516 burst fluence follows a powerlaw distribution ($\alpha$ = 2.0$\pm$0.2; see Figure~5) with the burst
duration. Similar correlations have been found for earthquakes with PL indices ranging from 1.4 to 1.8 (Gutenberg \& Richter 1956; Chen
et al. 1991) and for solar flares  with a PL index ranging from 1.53 to 1.73 (Crosby et al. 1993). Earthquake-like behaviour for SGRs were first pointed out by Cheng et al. (1996) using SGR 1806-20 with a PL index of 1.66.  The PL relation between the seismic moment ($\propto$ energy) and duration was shown by Lay \& Wallace (1995), which yielded an index of 3.03. Gogus et al. (2001) has also successfully investigated this correlation and obtained a PL index of 1.05$\pm$0.16 and 0.91$\pm$0.07 for SGR 1806-20 and SGR 1900+14, respectively. This behaviour shown by earthquakes was also interpreted on the basis of self-organized criticality (Bak et al. 1988), providing a framework for understanding complexity and scale invariance in systems showing irregular fluctuations. Self-organized criticality can be considered as a characteristic state of criticality formed by self-organization in a long transient period at the border of stability and chaos with the events following a powerlaw distribution. However, this theory cannot predict the strength or time of the next event (Bak et al. 1988). Moreover, we notice that this is the first time the PL index obtained from an SGR matches quite well with those measured in both earthquakes and solar flares, suggesting that SGR bursts are analogous to earthquakes and have both crustal and magnetospheric origins. 

We have also investigated the hardness-fluence and hardness-duration correlations for the 18 bursts observed from SGR 0501+4516. As suggested by other authors, hardness-duration and hardness-fluence anti-correlations are observed, as shown in Figures~6 (a) and (b), respectively. Moreover, as pointed out by Gogus et al. (2001), we see that the SGR bursts soften with increasing burst durations, the details of which will be presented in a follow-up paper. The hardness-fluence anti-correlation (see Figure~6 (b)) has been explained by Gogus et al. (2001) as due to the emitting plasma in local thermodynamic equilibrium with the radiative area decreasing at lower fluences, and secondly, due to the spectral intensity of the radiation field being below that of a BB, causing the emitting plasma temperature to remain in a narrow range and higher at lower luminosities. However, it is also suggested that these two possibilities depend on the rate of energy injection into the atmosphere (Gogus et al. 2001).

The cut-off energy $E_C$ obtained from the CPL model peaks around $\sim$~19.1$\pm$1.8 keV (Figure~4(b) and Table~4), consistent with that obtained using \textit{Konus-Wind} observations ($E_C$ in the range of 20-45 keV; Aptekar et al. 2009). The $\Gamma_{CPL}$ for SGR~0501+4516 is also in good agreement with those obtained for SGR 1900+14 and SGR 1806$-$20 (see Table~3 of Feroci et al. 2004 and Tables~9 \& 10 of Nakagawa et al. 2007). The magnetar model (Thompson \& Duncan 1996) suggests that the nonthermal emission observed in the SGRs originate from the hydromagnetic winds of particles in the magnetosphere and the cut-off energy $E_C$ can be associated with the plasma energy distribution in the magnetosphere. This contributes to the hard X-ray emission through particle acceleration leading to comptonization and particle bombardment of the surface (Thompson \& Duncan 1996; Mereghetti 2008) and/or from pairs created higher ($\sim$100 km) in the magnetosphere (Thompson \& Beloborodov 2005).


The spectra of SGR~0501+4516 were also well-fitted with a two-component BB model, with the low- and high-BB temperature clustered around 4.6$\pm$0.5 keV and 12.8$\pm$0.7 keV, respectively (Figure~4 and Table~4). These values are consistent with those reported by Enoto et al. (2009) using \textit{Suzaku} observations ($kT_{BB_{s}}$ = 3.3$^{+0.5}_{-0.4}$ keV and $kT_{BB_{h}}$ = 15.1$^{+2.5}_{-1.9}$ keV). The soft and hard BB component gave a Gaussian mean radii $R_{BB_{s}}$ = 7.2$\pm$1.5~$D_{10}$ km and $R_{BB_{h}}$ = 0.9$\pm$0.2~$D_{10}$~km, respectively. Several lines of interpretation were made to account for the BB+BB emission. Many recent studies have shown that the broadband (1--100 keV) spectroscopic studies of short SGR bursts are well approximated by the
sum of two blackbodies. Olive et al. (2004) reported that the time-integrated 2--150 keV energy spectrum of an intermediate flare from SGR~1900+14 observed with \textit{HETE-2} was best described by the sum of two BBs with temperatures $\sim$ 4.3~keV and $\sim$ 9.8~keV. These results were in line with those obtained for the short bursts from SGRs 1900+14  ($kT$ $\sim$ 3.23~keV and $\sim$ 9.65~keV) and 1806$-$20 ($kT$ $\sim$ 4~keV and $\sim$ 11~keV) by Feroci et al. (2004) and Nakagawa et al. (2007).  Similar results were also reported for the burst spectra of SGR~1900+14 and SGR~1627$-$41 with temperatures $kT_{BB_{s}}$ $\sim$ 3$-$5 keV and $kT_{BB_{h}}$ $\sim$ 9$-$10 keV (Israel et al. 2008; Esposito et al. 2008).  Olive et al. (2004) interpreted the high-BB component to be arising from a trapped fireball, with the high-BB temperature being consistent with the theoretically predicted temperature of a trapped fireball ($\sim$ 11 keV) and the low-temperature BB with the constant radius (almost 30 times larger than the average radius of the fireball) as originating from the surface of the neutron star. Furthermore, they suggest that the emission regions observed may be due to the radiation transfer effects in a superstrong magnetic field. In addition, these two BB temperatures can be considered as arising from the propagation of ordinary (O) and extraordinary (E) photons across the magnetosphere, with the photosphere of the E-mode photons located closer to the neutron star surface and the scattering photosphere of the O-mode photons higher up in the magnetosphere as suggested by Israel et al. (2008). Our results are also consistent with these interpretations based on the magnetar model.

In Figure~7, we have plotted the soft and hard component BB temperatures as a function of the square of their corresponding radii to investigate any possible correlations. The size of the emitting regions derived from the soft BB temperature, though does not appear to have a constant value, approximates the expected radius of the neutron star and a powerlaw fit to the data yields an index of $-$2.7$\pm$1.2. Similarly, we fitted a powerlaw to the hard BB temperature and their respective radii, and we obtain a PL index of $-$5.8$\pm$1.0. The Spearman correlation test gives the following anti-correlation results: $\rho$ = $-$0.5 ($P$ = 0.06) for the soft BB temperature and $\rho$ = $-$0.7 ($P$ = 0.01) for the hard BB temperature. This clearly demonstrates the existence of a strong anti-correlation between $kT_h$ and $R^2_{BB_{h}}$. Olive et al. (2004) found a nearly constant radius for the low-BB temperature, independent of the temperature whereas the high-BB temperature showed a clear anti-correlation between the radius and the temperature (see Figure~7 of Olive et al. 2004). Such a correlation ($R^2$ $\propto$ $kT^{-3}$) has been observed for SGR~1900+14 by Israel et al. (2008) during an intermediate outburst for the luminous phases of the flares ($L_{tot}$ $\ge$ 10$^{41}$ ergs~s$^{-1}$). Israel et al. (2008) identified the presence of a natural separation region around 25--30 km which corresponds to a critical surface where $B$ = $B_{QED}$, as predicted in the magnetar model (Duncan \& Thompson 1995). However, from our data on SGR 0501+4516, we notice that the minimum radius of the soft BB overlaps with the maximum radius of the hard BB component, thereby suggesting that a fraction of the E-mode photons can be locally converted to the O-mode by Compton scattering and by photon splitting if the effective temperature is high enough (Duncan \& Thompson 1995).

We have determined the soft and hard BB luminosities ($L_{BB_s}$ and $L_{BB_h}$), and plotted them against each other to explore their correlation. Figure~8 shows the bolometric luminosities of the two BBs for the bursts fitted with the BB+BB model. While Feroci et al. (1994) found a constant ratio for the bolometric luminosities of the bursts, we do not. However the quantities are well correlated with a correlation factor $\rho$ = 0.7 ($P$ = 6.1$\times$10$^{-3}$). We also fit the data with a powerlaw and obtain an index of 0.6$\pm$0.1, which is consistent with those obtained by Israel et al. (2008) for SGR 1900+14. The luminosities obtained for SGR 0501+4516 fall below $\sim$ 10$^{41}$ ergs s$^{-1}$, and hence, with the current statistics, we cannot make an inference on the saturation effect of the low BB luminosity as seen for SGR 1900+14 (Israel et al. 2008). The maximum bolometric luminosity obtained for the SGR 0501+4516 bursts is for the soft BB component with $L_{BB_s}$ = 4.1$\times$10$^{40}$ $D^2_{10}$~ergs s$^{-1}$ corresponding to the radius $R_{BB_s}$ = 14.2~$D_{10}$~km  and temperature $kT_{BB_s}$ = 6.7 keV. We compare this with the magnetic Eddington luminosity $L_{Edd, B}$ (Paczynski 1992; Thompson \& Duncan 1995) given by

\begin{equation}
L_{Edd, B} \sim 2\times10^{40}(\frac{B}{B_{QED}})^{4/3} (\frac{r}{R_{NS}})^{2/3} ~ergs~s^{-1}\\
\end{equation}

\noindent This gives a $B$-field value of $\sim$ 6.3$^{+1.0}_{-0.7}\times$10$^{13}$~G, which is consistent with the magnetic field strength of the dipolar component in the range $\sim$ 7$\times$10$^{13}$ $<$ $B_d$ $<$ 2$\times$10$^{14}$~G (Rea et al. 2009).


\section{Summary and Conclusions}
\label{6}

We have presented a quantitative analysis of the newly discovered SGR 0501+4516 observed by \textit{Swift}-BAT .  The spectra of the bursts from SGR 0501+4516 in the 15-150 keV energy range are best-fitted by both the CPL and BB+BB models. In the BB+BB model, the hotter temperature represents the smaller trapped fireball regions and the colder temperature corresponds to the regions consistent more or less with the radius of the neutron star. These temperatures are also  associated with the propagation of E-mode and O-mode photons across the photosphere as predicted in the magnetar model. The nonthermal emission observed from the short bursts seems to originate from the magnetosphere. All the above findings are consistent with the typical short bursts observed from other SGRs and with the magnetar model prediction of SGRs. However, we believe that more burst observations will help in further constraining the SGR emission mechanisms.

\acknowledgments
This research made use of NASA's Astrophysics Data System and of NASA's HEASARC maintained at the Goddard Space Flight Center.
 We acknowledge David Palmer, Kim Page and Paul O'Brien for useful conversations,
Craig Markwardt for his prompt response to queries on the data analysis, and Silvia Zane for her careful reading of the manuscript.
We acknowledge the MIT Kavli Institute of Astrophysics and Space Research where this
work was partly carried out. A. I acknowledges support from the Fulbright Commission. This research was supported by the Natural Sciences and Engineering Research Council of Canada and the Canada Research Chairs (CRC) program.

\newpage
\begin{figure*}[bh]
\includegraphics[width=0.5\textwidth]{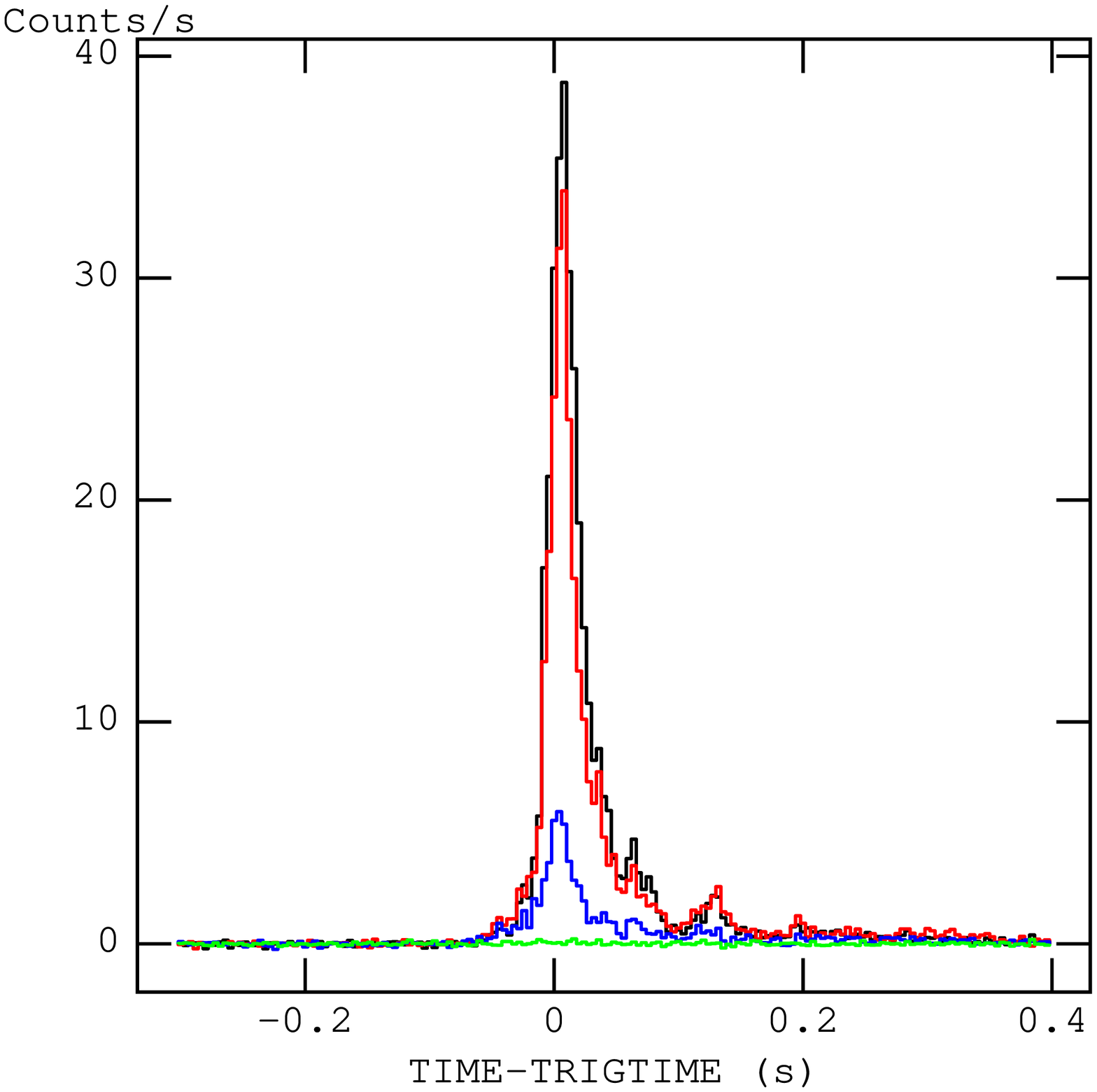}
\includegraphics[width=0.5\textwidth]{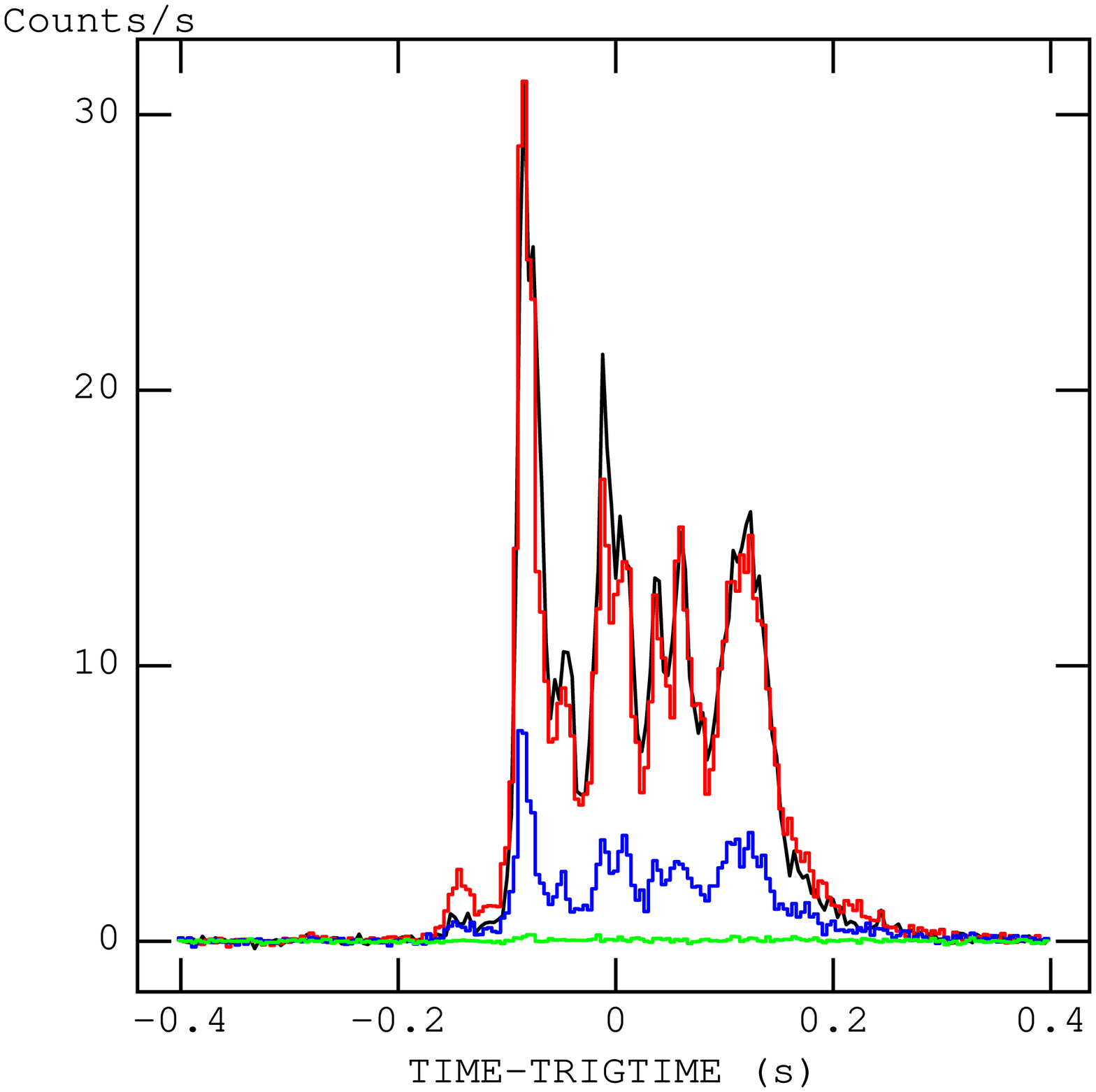}
\caption{Examples of the background-subtracted 4 ms lightcurves of SGR 0501+4516 bursts detected by \textit{Swift}-BAT in the 4 energy channels (Black: 15-25 keV, Red: 25-50 keV, Blue: 50-100 keV, Green: 100-150 keV) for Trigger numbers 00321481000 (Burst: 321481A; \textit{left}) and 00321551000 (Burst: 321551; \textit{right}).  }
\end{figure*}

\begin{figure}
\includegraphics[width=0.8\textwidth]{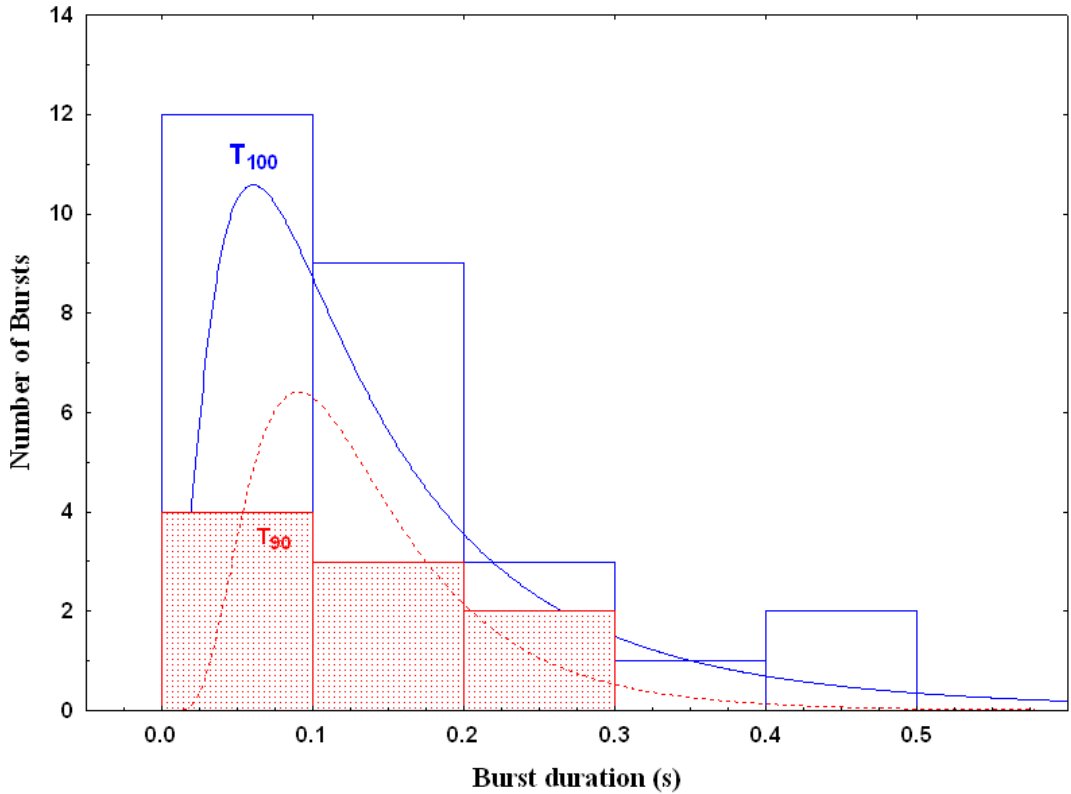}{\label{figure:T100}}
\caption{Distribution of the $T_{100}$ and $T_{90}$ durations of the SGR 0501+4516 bursts. The solid curve is obtained by fitting a
lognormal Gaussian model, which peaks at 0.06$\pm$0.02~s. Also shown by a dotted curve is the lognormal fit to the $T_{90}$ durations,
peaking at 0.09$\pm$0.02~s.}
\end{figure}

\begin{figure}
\subfigure[CPL spectrum]{\label{figure:CPL}\includegraphics[width=0.55\textwidth, angle=-90]{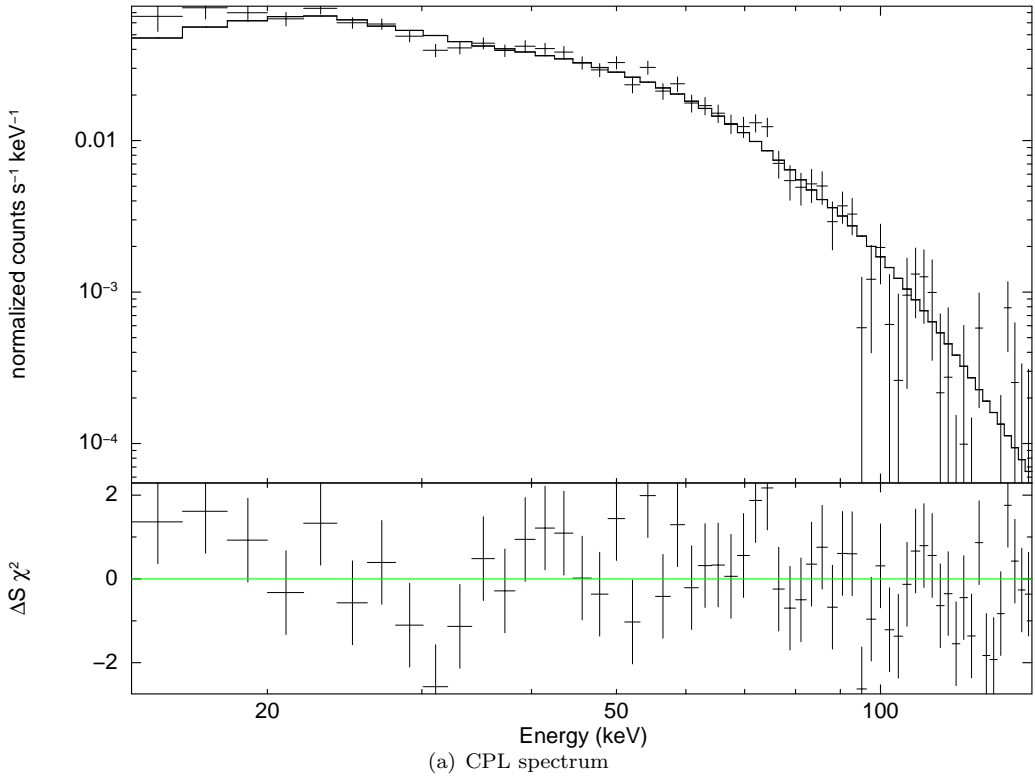}}
\subfigure[BB+BB spectrum]{\label{figure:BB+BB}\includegraphics[width=0.55\textwidth, angle=-90]{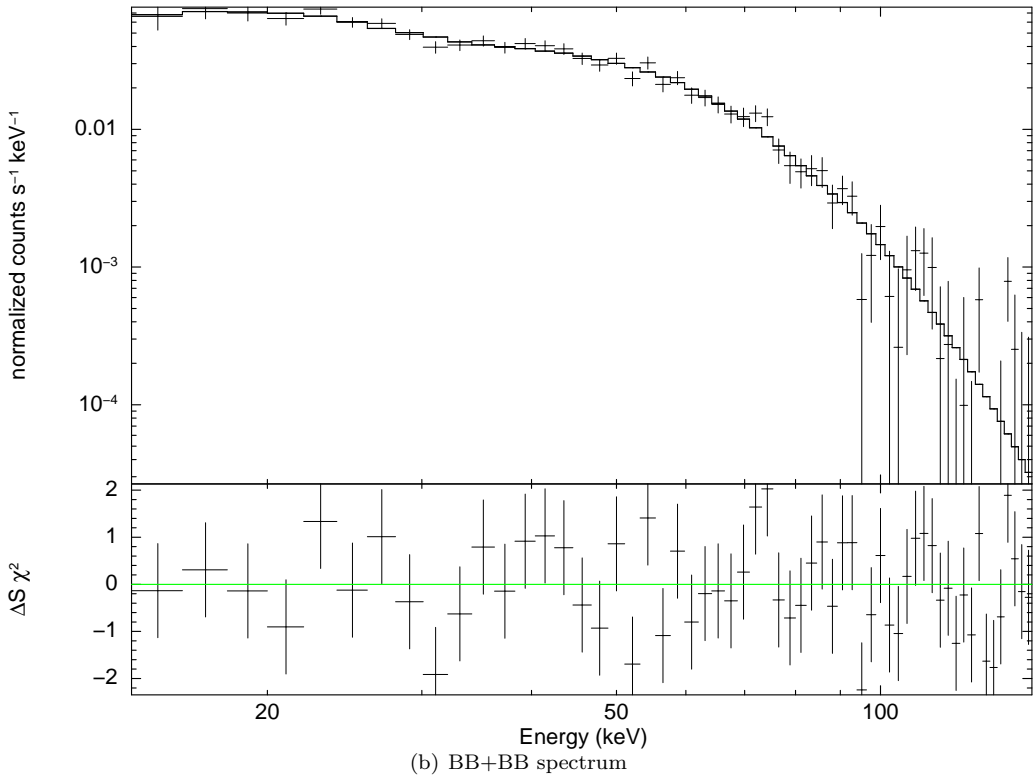}}
\caption{Sample spectra of an SGR 0501+4516 burst (321583E) of duration $T_{100}$ = 0.33~s. (a) CPL model fit  with $\langle\chi^2_{\nu}\rangle$ = 1.250 (56). (b) BB+BB model fit with $\langle\chi^2_{\nu}\rangle$ = 1.009 (55). }
\end{figure}

\begin{figure}
\includegraphics[width=0.5\textwidth]{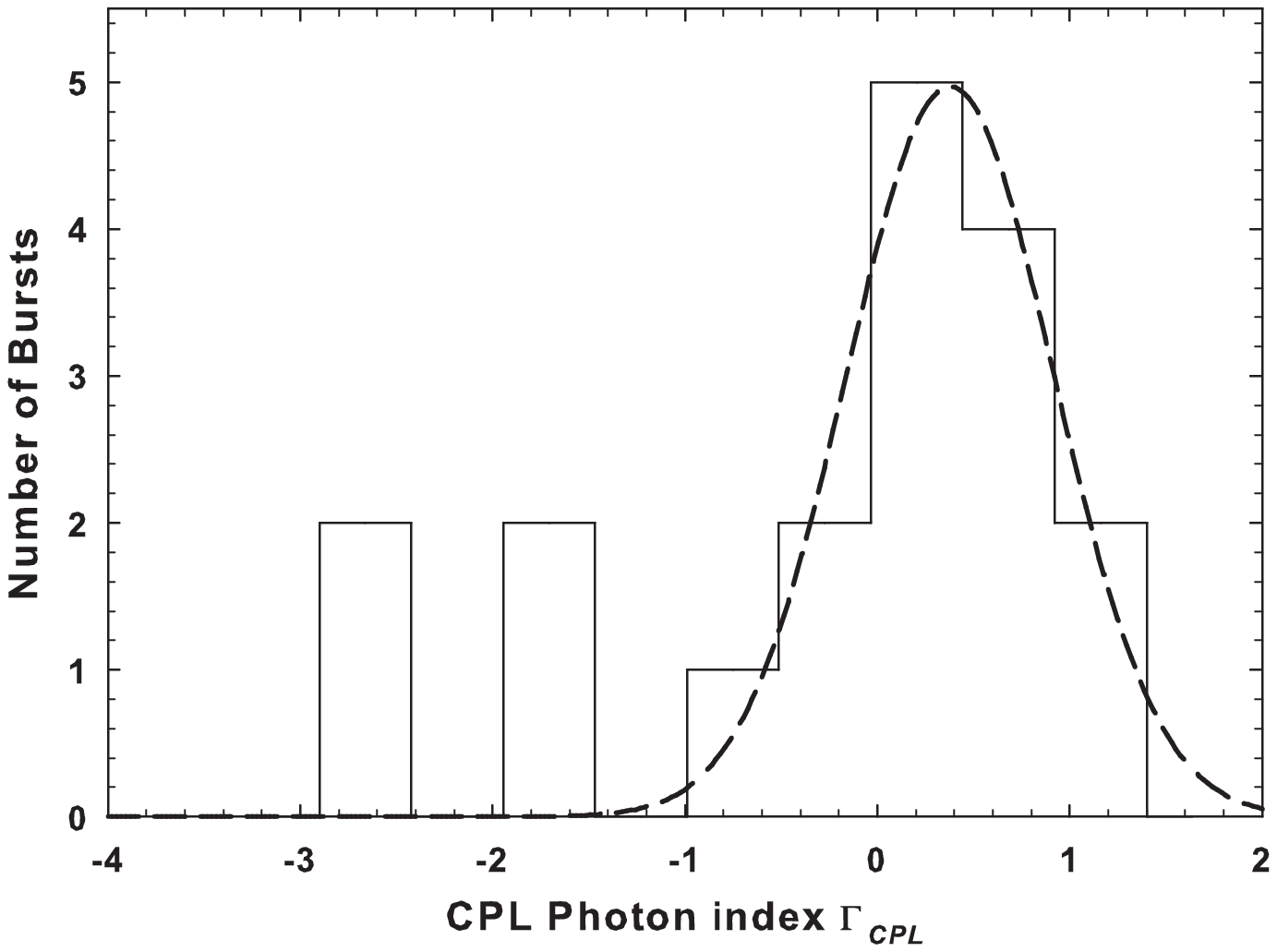}
\includegraphics[width=0.5\textwidth]{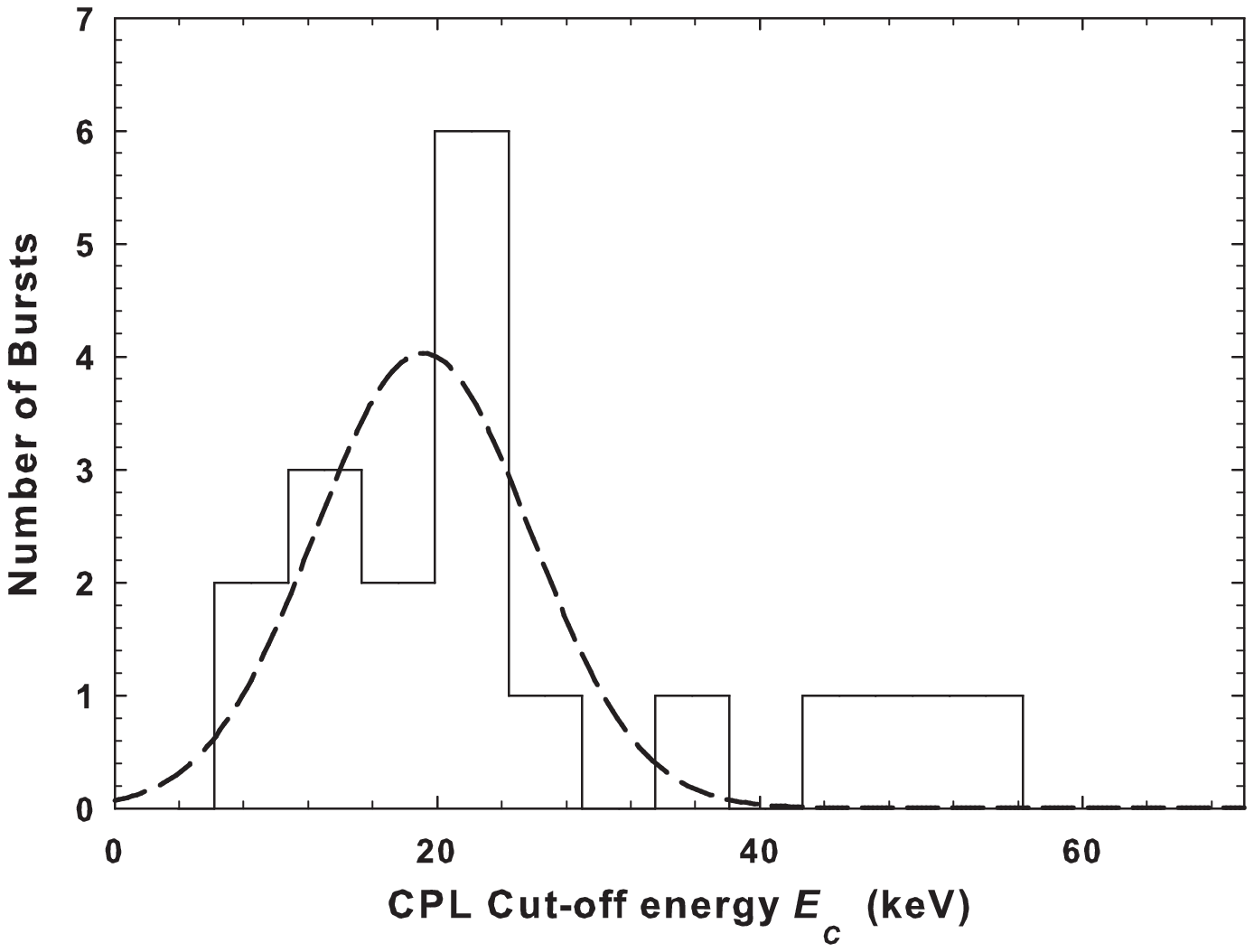}
\includegraphics[width=0.5\textwidth]{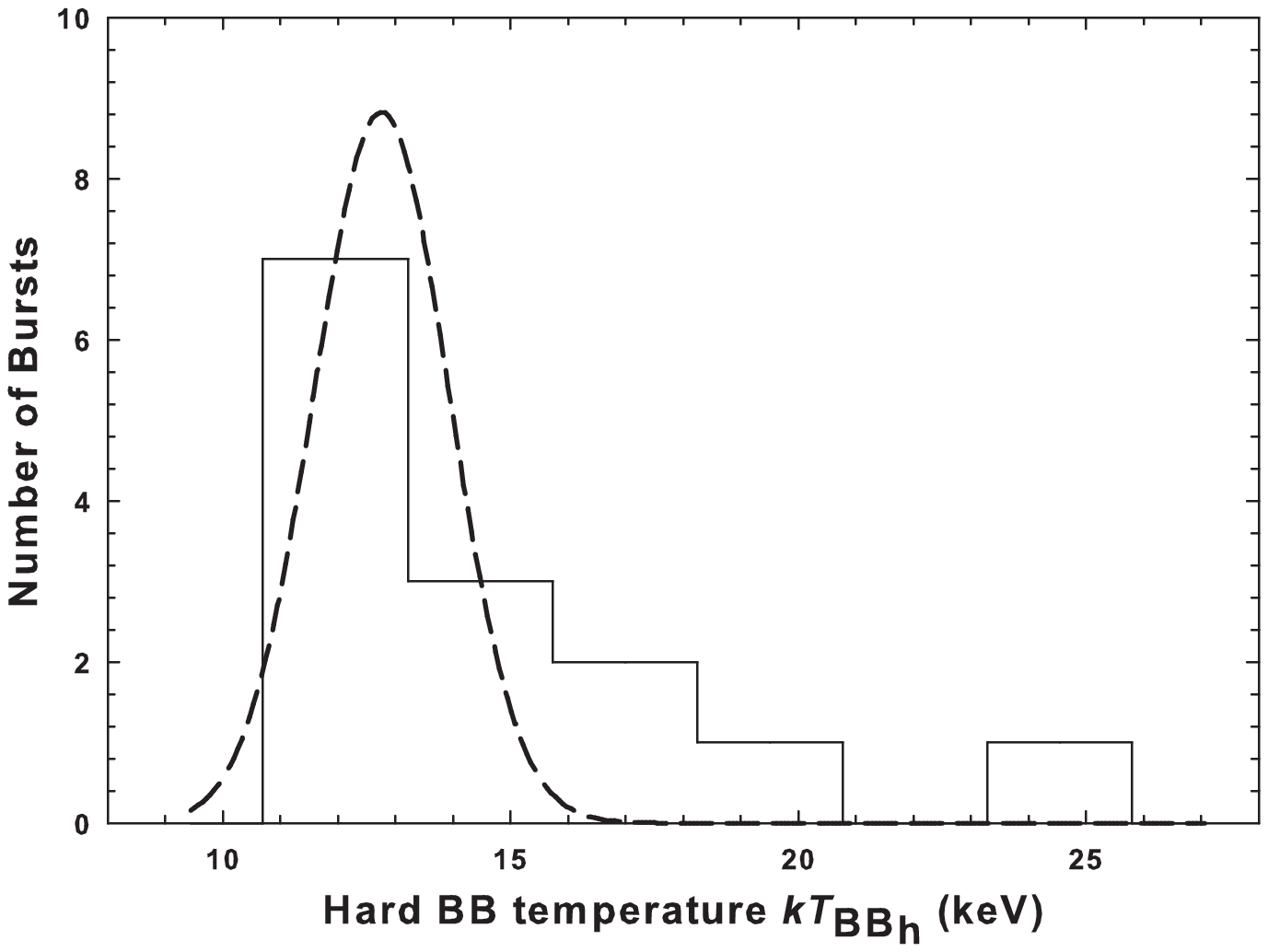}
\includegraphics[width=0.5\textwidth]{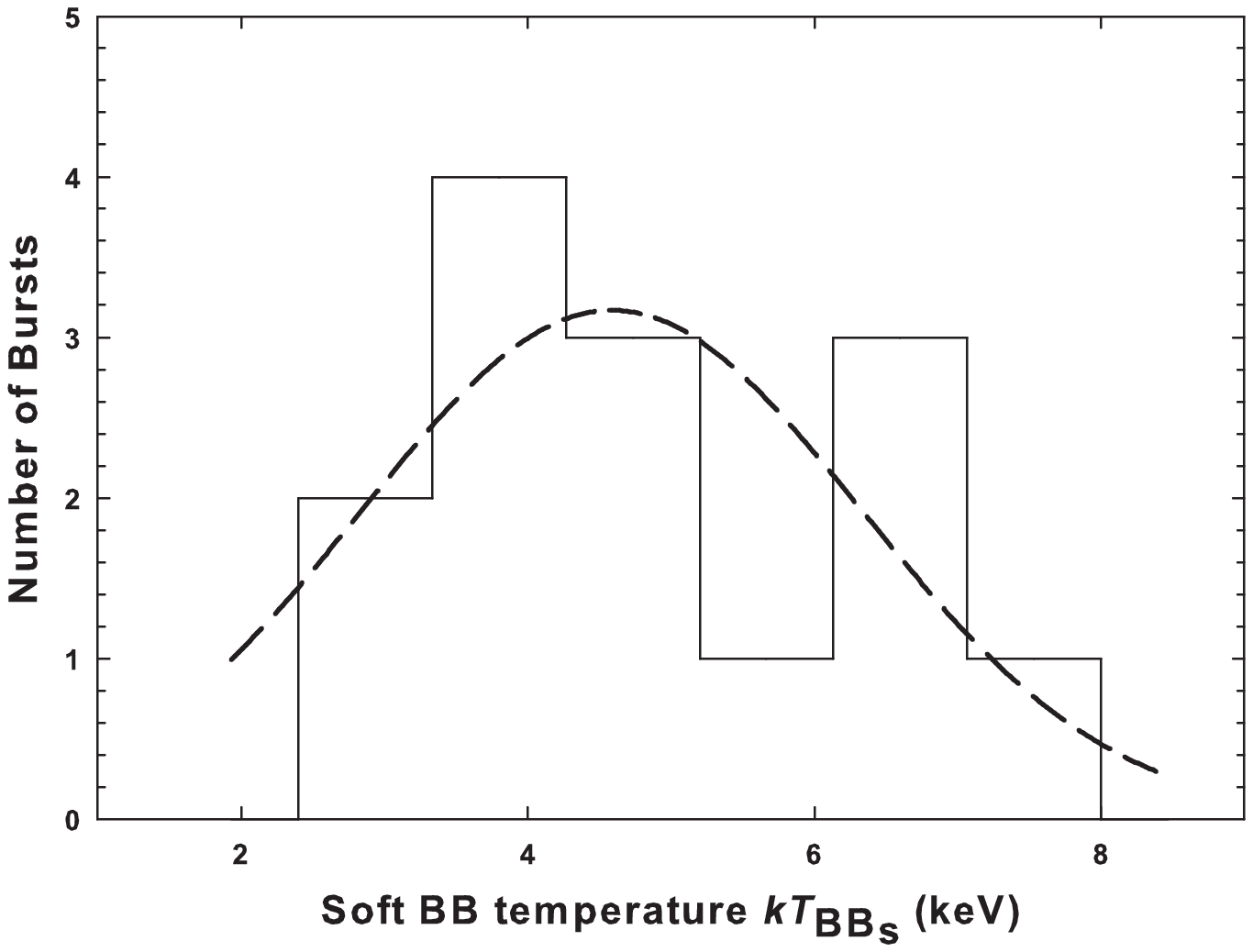}
\includegraphics[width=0.5\textwidth]{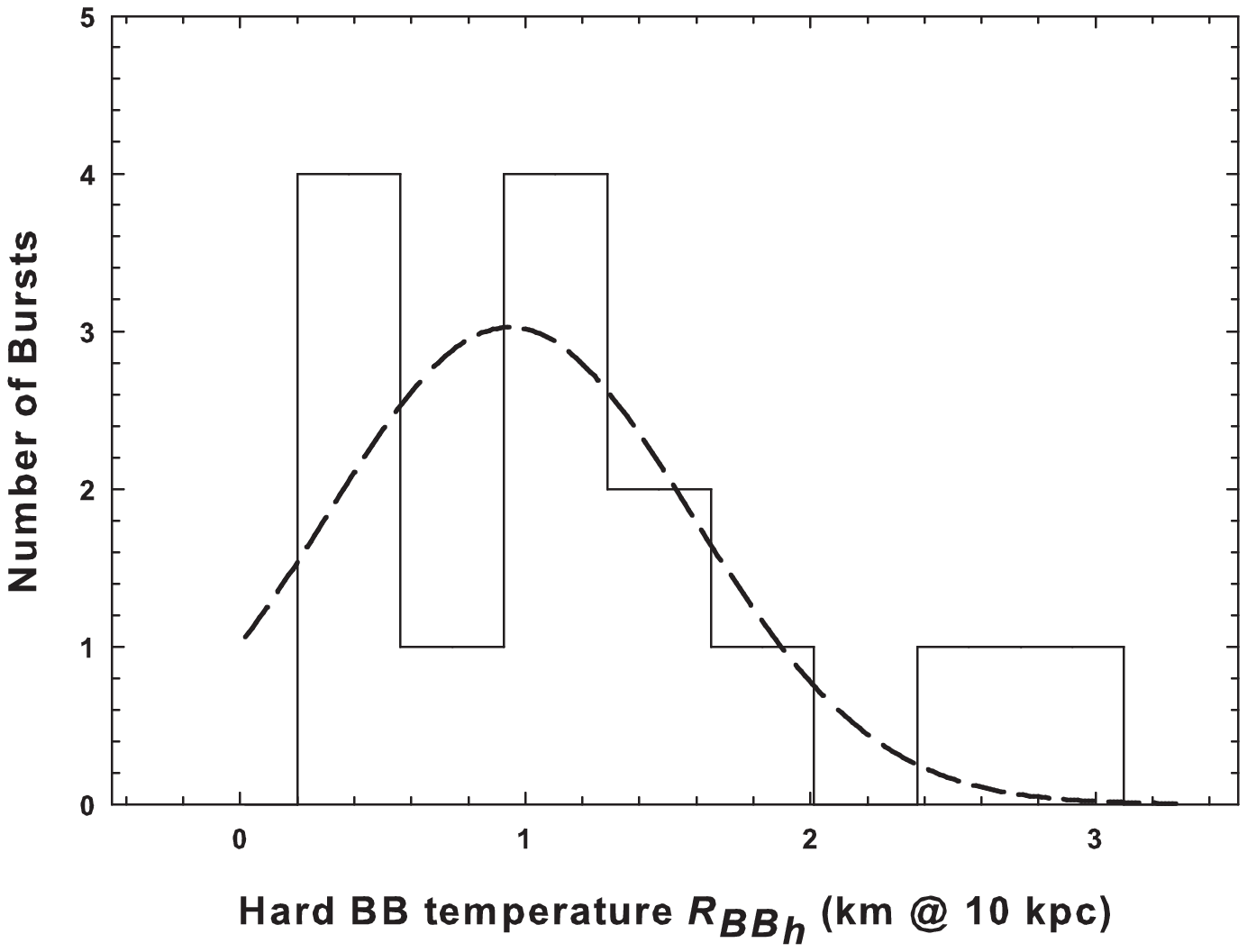}
\includegraphics[width=0.5\textwidth]{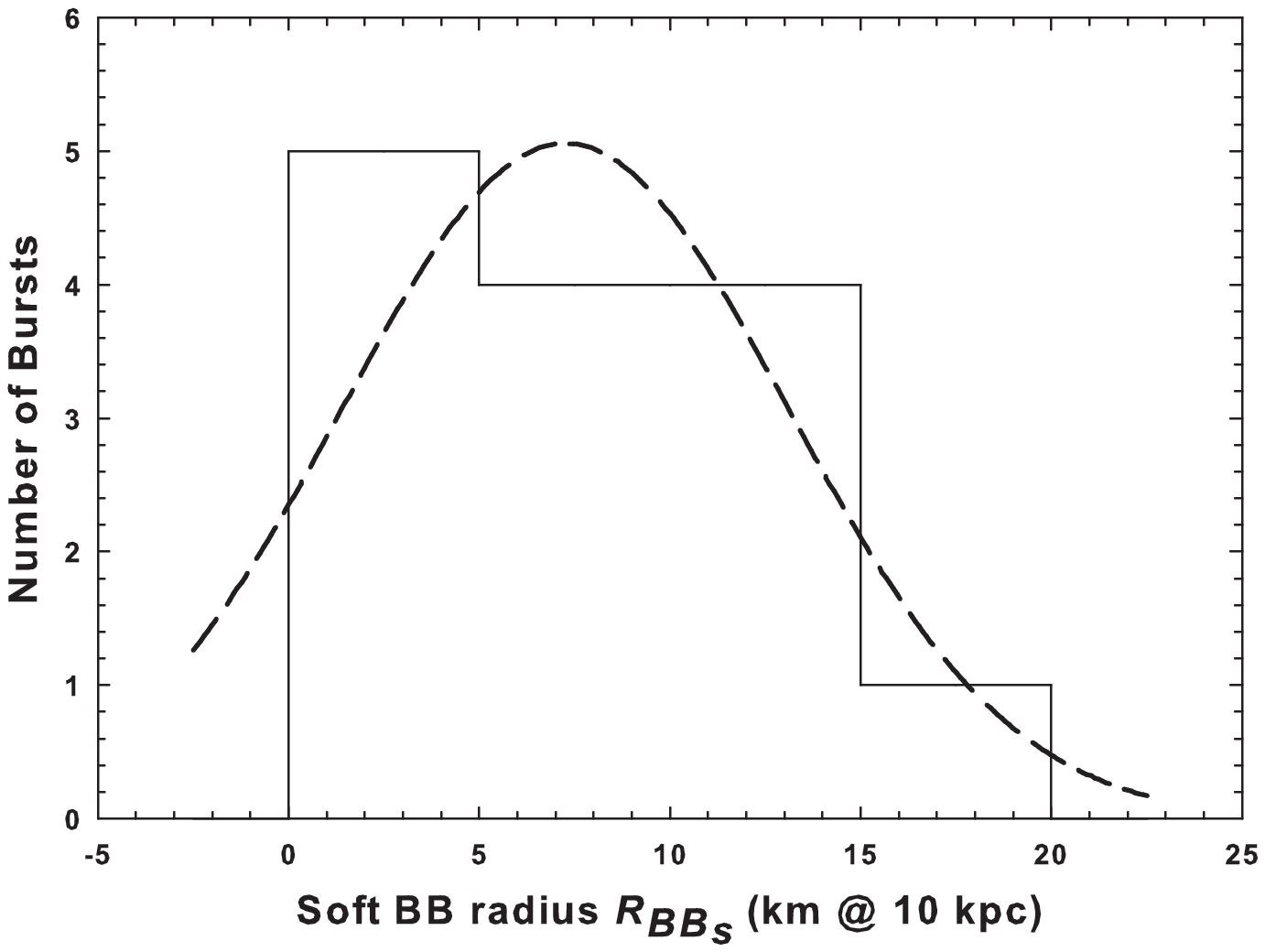}
\caption{Histogram plots for the best-fit spectral parameters fitted with a Gaussian. The spectral indices ($\Gamma_{CPL}$), and cut-off energy $E_C$ are obtained from the CPL model. The soft BB temperature $kT_{BB_{s}}$, hard BB temperature $kT_{BB_{h}}$, and the inferred soft and hard BB radii (R$_{BB_{s}}$, R$_{BB_{h}}$) are obtained from the BB+BB model. Details of the Gaussian fit are given in Table~4.}
\end{figure}

\begin{figure}
\includegraphics[width=\textwidth]{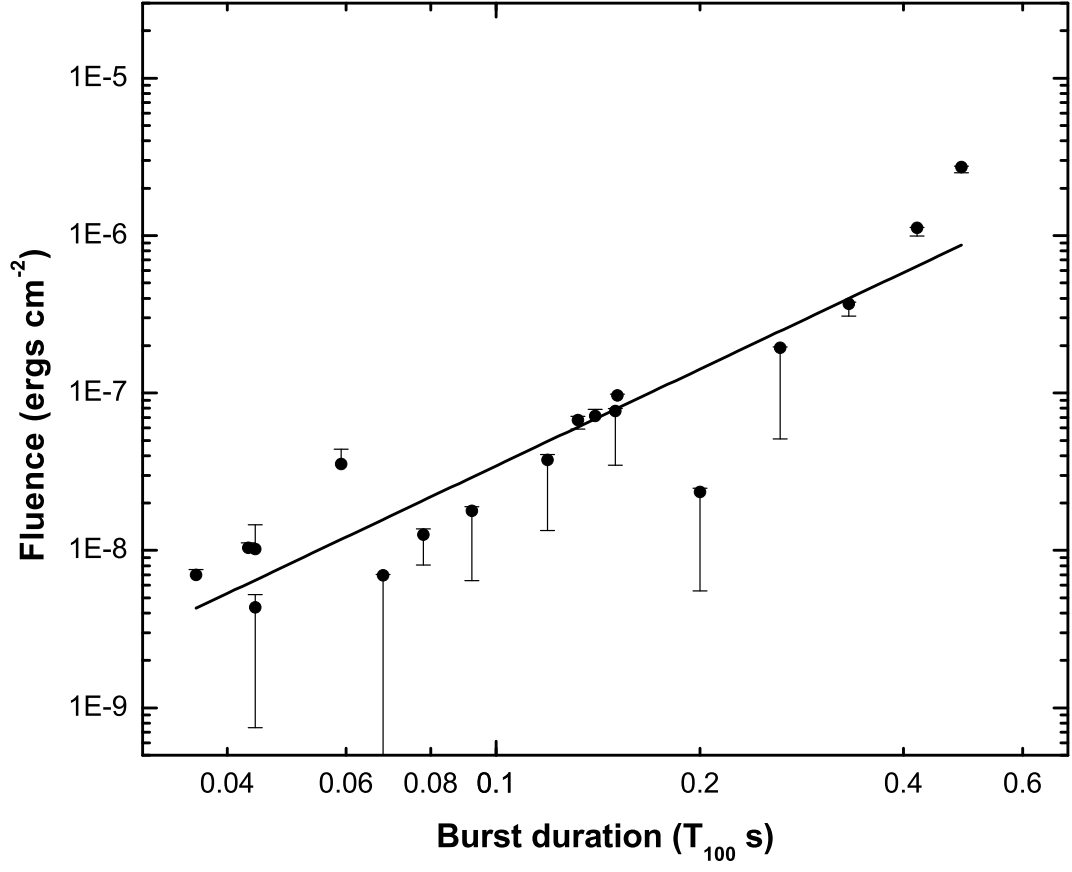}
\caption{Correlation plot for SGR 0501+4516 burst fluence vs. $T_{100}$ fitted with a PL of index 2.0$\pm$0.2 ($\rho$ = 0.9, $P$ = 3.0$\times$10$^{-7}$). }
\end{figure}

\begin{figure}
\subfigure[HR-Duration]{\label{figure:HR-T100}\includegraphics[width=0.75\textwidth]{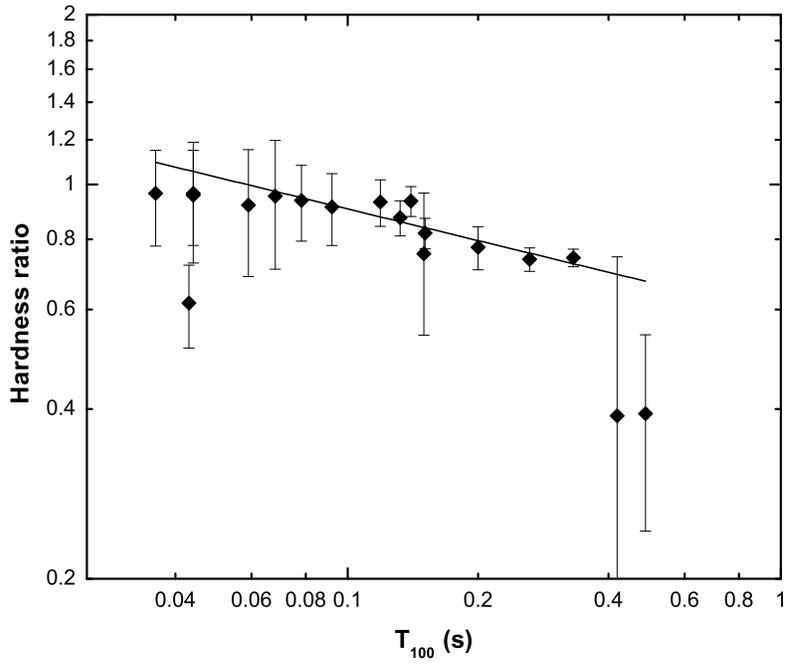}}
\subfigure[HR-Fluence]{\label{figure:HR-Fluence}\includegraphics[width=0.75\textwidth]{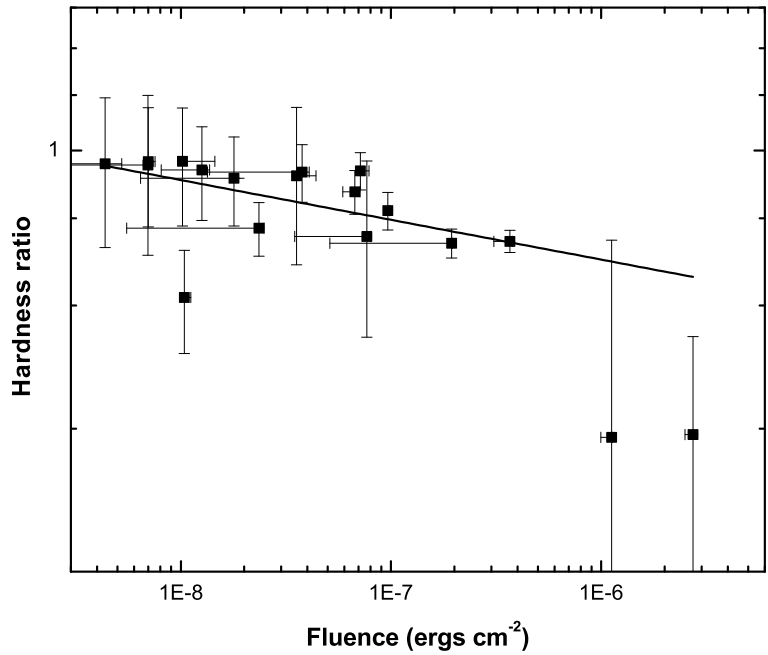}}
\caption{(a) Hardness ratio vs. $T_{100}$ fitted with a PL of index $-$0.20$\pm$0.03 ($\rho$ = $-$0.7, $P$ = 6.5$\times$10$^{-4}$). (b) Hardness ratio vs. fluence fitted with a PL of index $-$0.10$\pm$0.02 ($\rho$ = $-$0.8, $P$ = 1.8$\times$10$^{-4}$).}
 \end{figure}

\begin{figure}
\includegraphics[width=0.75\textwidth]{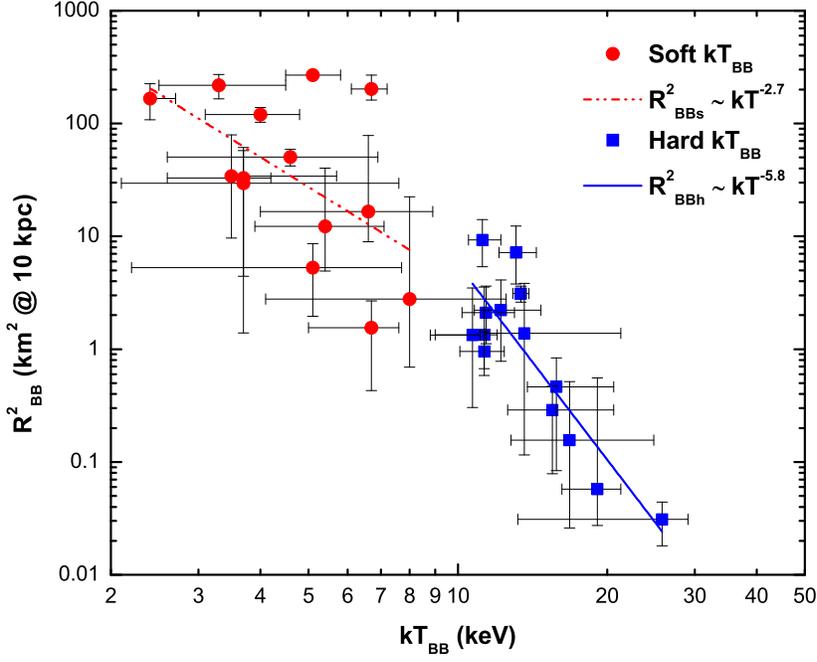}
\caption{Square of the radii of the BB+BB model components as a function of their temperatures.
The red circles and blue squares denote the soft and hard $kT$ values, respectively. The soft $kT$ values show an anti-correlation of $\rho$ = $-$0.5, $P$ = 0.06 with $R^2_{BB_{s}}$, and the hard $kT$ shows a strong anti-correlation of $\rho$ = $-$0.7, $P$ = 0.01 with $R^2_{BB_{h}}$.}
\end{figure}

\begin{figure}
\includegraphics[width=0.75\textwidth]{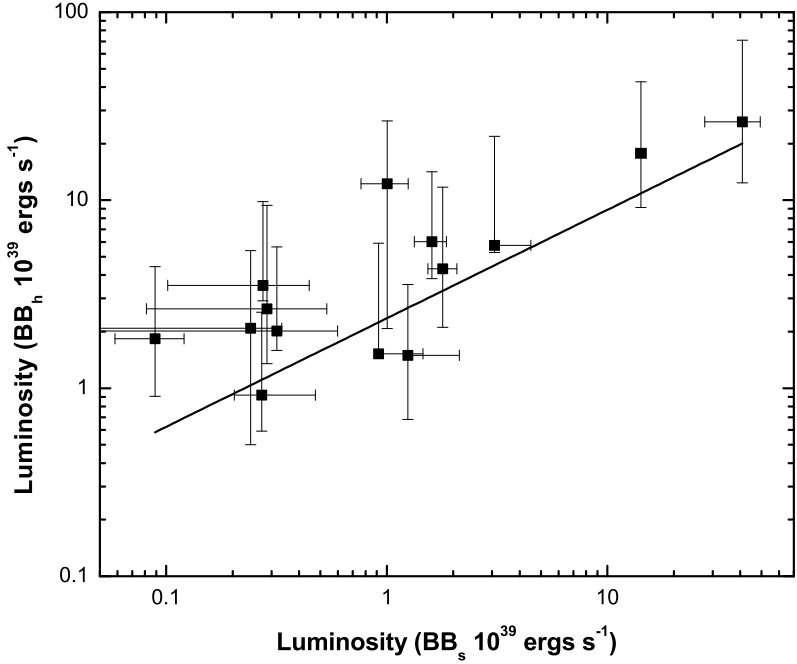}
\caption{Bolometric luminosity of the hard BB component ($L_{BB_h}$) vs. that of the soft BB component ($L_{BB_s}$) in units of 10$^{39}$ ergs~s$^{-1}$. The solid line is the powerlaw fit which gives the relation $L_{BB_s}$ = ($L_{BB_h}$)$^{\alpha}$ where $\alpha$ = 0.6$\pm$0.1.}
\end{figure}

\end{document}